\documentstyle[12pt]{article}
\newfont{\ffont}{msym10}                        
\newcommand{\beq}{\begin{equation}}             
\newcommand{\eeq}{\end{equation}}               
\newcommand{\bqry}{\begin{eqnarray}}            
\newcommand{\eqry}{\end{eqnarray}}              
\newcommand{\bqryn}{\begin{eqnarray*}}          
\newcommand{\eqryn}{\end{eqnarray*}}            
\newcommand{\NL}{\nonumber \\}                  
\newcommand{\preprint}[1]{\begin{table}[t]      
            \begin{flushright}                  
            \begin{large}{#1}\end{large}        
            \end{flushright}                    
            \end{table}}                        
\newcommand{\PD}[2]                             
    {\frac{\partial^{#2}}{\partial #1^{#2}}}    
\catcode`\@=11 \@addtoreset{equation}{section} 
\renewcommand{\theequation}                    
         {\arabic{section}.\arabic{equation}}  
\begin{document}
\preprint{TAUP-2161-94 \\  }
\title{Relativistic Mass Distribution \\ in Event--Anti-event System \\
and \\ ``Realistic'' Equation of State \\ for Hot Hadronic Matter}
\author{\\ L. Burakovsky\thanks {Bitnet: BURAKOV@TAUNIVM.TAU.AC.IL.} \
and L.P. Horwitz\thanks
  {Bitnet: HORWITZ@TAUNIVM.TAU.AC.IL. Also at Department of Physics,
  Bar-Ilan University, Ramat-Gan, Israel  } \\ \ }
\date{School of Physics and Astronomy \\ Raymond and Beverly Sackler
Faculty of Exact Sciences \\ Tel-Aviv University,
Tel-Aviv 69978, ISRAEL}
\maketitle
\begin{abstract}
We find the equation of state $p,\rho \propto T^6,$ which gives the value
of the sound velocity $c^2=0.20,$ in agreement with the ``realistic''
equation of state for hot hadronic matter suggested by Shuryak, in the
framework of a covariant relativistic statistical mechanics of an
event--anti-event system with small chemical and mass potentials. The
relativistic mass distribution for such a system is obtained and shown
to be a good candidate for fitting hadronic resonances, in agreement
with the phenomenological models of Hagedorn, Shuryak, {\it et al.}
This distribution provides a correction to the value of specific heat
3/2, of the order of 5.5\%, at low temperatures.
\end{abstract}

\bigskip
\section{Introduction}
One of the main goals of experiments with high-energy nuclear collisions
is to produce and to study hadronic matter, in particular, trying to
reach conditions at which the phase transitions into the quark-gluon
plasma phase can take place [1]-[5]. The physics of hot hadronic matter
has not been studied much, although such matter is already produced in
current experiments in high-energy physics.

The experimental data on multiple hadron production obtained in recent
years are in agreement with the main consequences of the theory
formulated by Landau \cite{Lan} over 40 years ago. However, with a
sufficiently quantitative approach, it becomes necessary to consider a
number of physical effects which bring about certain modifications of the
results obtained in the fundamental work [6]. For instance, the solution
of the equations of motion obtained by Landau differs from a numerical
calculation of Melekhin \cite{Mel1}, as a result of an inaccurate
estimate. A more accurate analytic solution was given in \cite{Shu1}.

The equation of state in Landau's work is taken to be $p=\rho /3 $ (where
$p$ is the pressure and $\rho $ is the energy density), corresponding to
an ultra-relativistic gas. However, the lab energies $s\equiv (p_1+p_2)^
2\stackrel{<}{\sim }3\times 10^3$ GeV$^2$ correspond to initial
temperatures $T\stackrel{<}{\sim }1$ GeV. In this temperature range,
the interaction of the hadrons is strong and has mainly a resonant
character, the masses of the resonances being comparable with the
temperature. Thus, hadronic matter under these conditions is neither
an ideal nor an ultra-relativistic gas.

Corrections in the equation of state due to the interaction of the
hadrons have been discussed in the literature for the last three
decades [1],[4],[8]-[14]. These considerations were based mostly on a
phenomenological model in which the Landau theory is applied to all
particles except the leading ones, i.e., the fragments of the initial
particles, whose characteristics are taken directly from experimental
data. The framework for the latter considerations is the QCD phase
transition of hadronic matter into the quark-gluon plasma.

Hot hadronic matter, in which volumes per particle are a few cubic
fermis, is certainly made out of individual hadrons. It is clear that,
at low temperatures $T<<m_\pi ,$ one has a very rare (and therefore
ideal) gas of the lightest hadrons, the pions. As the temperature is
raised and the gas becomes more dense, one should take into account
interactions among the particles. This was done using the following three
approaches: (i) the low-temperature expansion, (ii) the resonance gas,
(iii) the quasi-particle gas.

The first approach for the pion gas is based on the Weinberg theory of
pion interactions \cite{Wei}, which uses the non-linear Lagrangian
containing all processes quadratic in pion momentum. Its first
application to calculation of the thermodynamic parameters of the pion
gas was done in
ref. [1]. One of the important consequences of this work was
that, after isospin averaging, all corrections quadratic in momenta
cancel each other, and corrections proportional to the pion mass
(Weinberg $\pi \pi $ scattering lengths) are nearly compensated,
producing negligible corrections at the 1\% level. Corrections of the
second order in the Weinberg Lagrangian were also estimated in [1], and
a much more systematic study of the problem including quartic terms in
the mesonic Lagrangian were made in ref. \cite{GL}. Without going into
discussion of these works we remark that the applicability of this
approach is limited by temperatures $T<100$ MeV, for which typical
collision energies are significantly below resonance .However, such $T$
are lower than even the lowest temperature available in experiments,
because the so-called break-up temperatures are typically $T\simeq 120-
150$ MeV.

The idea of resonance gas was first suggested by Belenky and Landau
\cite{BL}, as early as 1956. They used the Beth-Uhlenbeck formula
\cite{BU}, well known in the theory of the non-ideal gases \cite{LL},
which relates the second virial coefficient with scattering phase shifts
$\delta _l(p).$ In particular, in this approximation the ``internal''
part of the statistical sum can be written as a sum over bound states
plus the scattering part\footnote{This formula is written down for the
nonrelativistic case, as it appears in textbooks on statistical
mechanics, e.g., in \cite{LL}.}:
\beq
Z_{{\rm int}}=\sum _ne^{-E_n/T}+\frac{1}{\pi }\sum _l\int dp\frac{
d\delta _l(p)}{dp}e^{-p^2/2mT}.
\eeq
Using Eq. (1.1) one can prove that a narrow and elastic resonance
is analogous to adding one extra physical state to the system.

This observation was later used by many authors, in particular, by
Hagedorn \cite{Hag}, who has generalized this statement to the idea that
all hadrons (both stable and resonances) should be treated as real
degrees of freedom of hadronic matter.The statistical bootstrap model
\cite{Hag},\cite{Fra} initiated by Hagedorn, which requires
that the spectra of the  resonances and of the system as a whole
coincide, provides a resonance spectrum
\beq
\rho (m)\sim m^a\exp (m/T_0),
\eeq
where $a$ and $T_0$ are some parameters. The statistical sum diverges
when $T>T_0,$ which indicates that the theory involves a limiting
temperature (the so-called ``hadronic boiling point'' \cite{Hag})
to which the system can be heated. Such behavior indicates that there is
some phase transition, and that the language used is meaningless above
it. This phase transition, as well as thermodynamic consequences of the
mass spectrum (1.2), were discussed in ref. \cite{Car}. The expression
(1.2) with $a=-5/2$ is in good agreement with experiment in the
resonance mass region $m\stackrel{<}{\sim }1.2$ GeV \cite{Shu1}.

In 1972, Shuryak \cite{Shu1}, in place of (1.2), used the simple power
parametrization
\beq
\rho (m)\sim m^k,
\eeq
which also describes experiment in this mass region for $k\approx 3.$
The energy density is expressed by \cite{Shu1} $(E^2={\bf p}^2+m^2)$
\beq
\rho =\sum _\sigma \int dm\rho (m)\int \frac{d^3{\bf p}}{(2\pi )^3}
\frac{E}{\exp (\frac{E}{T})+\sigma },\;\;\;\sigma =\pm 1.
\eeq
Since $\rho (m)$ is a rapidly growing function, the main contribution to
integrals of type (1.4) is given by the mass region in which $\exp (m/T)
>>1,$ and the difference in properties of bosons and fermions is
irrelevant. Upon substitution of (1.3) into (1.4) one finds that the
energy density and, analogously, the pressure, defined by
\beq
p=\frac{1}{3}\sum _\sigma \int dm\rho (m)\int \frac{d^3{\bf p}}{(2\pi
)^3}\frac{{\bf p}^2}{E}\frac{1}{\exp (\frac{E}{T})+\sigma },
\eeq
is proportional to $T^{k+5}.$ Therefore, the velocity of sound $c,$
defined by
\beq
c^2=\frac{dp}{d\rho },
\eeq
turns out to be a temperature independent constant, and the equation of
state belongs to the class considered by Melekhin \cite{Mel2}. For this
case, the expressions for energy and pressure are as follows,
\beq
\rho =\lambda T^{k+5},\;\;\;p=\frac{\lambda }{k+4}T^{k+5},
\eeq
where $\lambda $ is a certain constant. Consequently, for $k\approx 3$
[8]
\beq
c^2=\frac{1}{k+4}\approx 0.14.
\eeq
This numerical value of $c^2,$ although less than the ideal gas value
$c^2=1/3,$ is obtained very approximately due to replacement of the sum
over the resonances by the integral (1.4). Strange and baryonic
resonances may only occur in pairs, which is not taken into account in
derivation of (1.8). Moreover, at temperatures $T\stackrel{>}{\sim }m,$
which is the case we are interested in, quantum field theory requires the
inclusion of antiparticles which becomes important and therefore should
be taken into account.

In 1975, Zhirov and Shuryak \cite{ZS} used the Beth-Uhlenbeck
method \cite{BU}, which reduces the problem in the case of a purely
resonant interaction to a simple case of a gas consisting of a mixture of
stable and unstable particle (resonances) treated on an equal footing.
Their calculations give the value of the velocity of sound
$c^2=0.18-0.21,$ for the range of temperatures $0.2-1.0$ GeV, which is
appreciably smaller than than the asymptotic value $c^2=1/3$ and larger
than the estimate $c^2\approx 0.14$ of ref. \cite{Shu1}.

In his book \cite{QCD} of 1988, Shuryak proposed the ``realistic''
equation of state (referred to also in [1])
\beq
p\simeq \left(20{\rm GeV}^{-2}\right)T^6,
\eeq
describing the behavior of hadronic matter in the temperature range
$0.2{\rm GeV}\stackrel{<}{\sim }T\stackrel{<}{\sim }1.0{\rm GeV}$ (which
gives $c^2=0.20),$ in good agreement with the relevant experimental data.

A similar equation was obtained in refs. \cite{SR},\cite{ME} within the
concept of hadronic-plasma phase transitions, for a kind of chiral
bag model for the quark-gluon plasma cluster.

In ref. \cite{Shu2} Shuryak, in order to describe the behavior of the
energy density with temperature $\rho \propto T^6,$ proposed to consider
hot hadronic matter as a gas of quasiparticles, which have quantum
numbers of the original mesons, but with dispersion relations
modified by the interaction with matter (similarly to Landau's idea
of ``rotons'', to explain the growth of the energy density of liquid
$^4He$ with temperature more rapidly than $T^4).$
Since in such an approach hadrons are included as physical
degrees of freedom of hot hadronic matter, they are assumed not to be
absorbed too strongly, i.e., to be good quasiparticles. In ref.
\cite{Shu2} an attempt was made to guess what these dispersion relations
should be, in order to obtain the expected behavior of the energy density
with temperature. In \cite{Shu4} these dispersion relations were
calculated in explicit form.

In this paper we show that the same results can be obtained within the
framework of a manifestly covariant relativistic statistical mechanics
[24]-[27]. We shall review this framework briefly in the next section.
First a possible correlation of the manifestly covariant theory with the
hadronic spectrum of the Hagedorn-Frautschi form was discussed by Miller
and Suhonen [28]. They remarked that an approach based on the grand
canonical distribution function of ref. [24] gives results similar to
the ones for the resonance gas in the case $\mu _K\simeq 0,$ where $\mu _
K$ is the additional mass potential of the ensemble [24]. In the present
paper we study a free relativistic gas in the limiting case\footnote{For
finite temperature this case is realized by $\mu _K\rightarrow 0.$}
$\frac{M}{\mu _KT}\rightarrow \infty $ (where $M$ is an intrinsic scale
parameter for the motion in space-time [29]), taking account of
antiparticles. For such a particle-antiparticle system with $\mu \simeq 0
$ (i.e., zero net particle charge) we find the equation of state
$p,\rho \propto T^6,$ which gives the value of the velocity of sound
$c^2=0.20.$ We obtain the relativistic mass distribution for this system
and show that this distribution is a good candidate for fitting the
hadronic resonances, in agreement with the theory of Hagedorn [9],[10].

\section{Relativistic $N$-body system}
In the framework of a manifestly covariant relativistic statistical
mechanics, the dynamical
evolution of a system of $N$ particles, for the classical case, is
governed by equations of motion that are of the form of Hamilton
equations for the motion of $N$ $events$ which generate the space-time
trajectories (particle world lines) as functions of a continuous
Poincar\'{e}-invariant parameter $\tau ,$ called the historical time
[29],[30]. These events are characterized by their positions
$q^\mu =(t,{\bf q})$ and energy-momenta $p^\mu =(E,{\bf p})$ in an
$8N$-dimensional phase-space. For the quantum case, the system is
characterized
by the wave function $\psi _\tau (q_1,q_2,\ldots ,q_N)\in L^2
(R^{4N}),$ with the measure $d^4q_1d^4q_2\cdots d^4q_N\equiv d^{4N}q,$
$(q_i\equiv q_i^\mu ;\;\;\mu =0,1,2,3;\;\;i=1,2,\ldots ,N),$ describing
the distribution of events, which evolves with a generalized
Schr\"{o}dinger equation [29]. The collection of events (called
``concatenation'' [31]) along each world line corresponds to a
{\it particle,} and hence, the evolution of the state of the $N$-event
system describes, {\it a posteriori,} the history in space and time of
an $N$-particle system.

For a system of $N$ interacting events (and hence, particles) one takes
[31]
\beq
K=\sum _i\frac{p_i^\mu p_{i\mu }}{2M}+V(q_1,q_2,\ldots ,q_N),
\eeq
where $M$ is a given fixed parameter (an intrinsic property of the
particles), with the dimension of mass, taken to be the same for all the
particles of the system. The Hamilton equations are
$$\frac{dq_i^\mu }{d\tau }=\frac{\partial K}{\partial p_{i\mu }}=\frac{p_
i^\mu }{M},$$
\beq
\frac{dp_i^\mu }{d\tau }=-\frac{\partial K}{\partial q_{i\mu }}=-\frac{
\partial V}{\partial q_{i\mu }}.
\eeq
In the quantum theory, the generalized Schr\"{o}dinger equation
\beq
i\frac{\partial }{\partial \tau }\psi _\tau (q_1,q_2,\ldots ,q_N)=K
\psi _\tau (q_1,q_2,\ldots ,q_N)
\eeq
describes the evolution of the $N$-body wave function
$\psi _\tau (q_1,q_2,\ldots ,q_N).$

\subsection{Ideal relativistic identical system}
To describe an ideal gas of events obeying Bose-Einstein/Fermi-Dirac
statistics in the grand canonical ensemble, we use the expression for
the number of events found in [24],
\beq
N=\sum _{k^\mu }n_{k^\mu }=
\sum _{k^\mu }\frac{1}{e^{(E-\mu +\mu _K\frac{m^2}{2M})/T}\mp 1},
\eeq
where $\mu _K$ is the additional mass potential [24], and $m^2\equiv -k^2
\equiv -k^\mu k_\mu .$ Replacing the sum over $k^\mu $ by an integral,
one obtains for the density of events per unit space-time volume $n
\equiv N/V^{(4)}$ [27],
\beq
n=\frac{1}{4\pi ^3}\int _0^\infty \frac{m^3\;dm\;\sinh ^2\beta \;d\beta }
{e^{(m\cosh \beta -\mu +\mu _K\frac{m^2}{2M})/T}\mp 1}.
\eeq
The integration results in the following expression for $n$ [27]:
\beq
n=\frac{1}{(2\pi )^3}\frac{M^2}{\mu _K^2}T^2\sum _{s=1}^\infty (\pm 1)^{
s+1}\frac{e^{\frac{s\mu }{T}}}{s^2}\Psi (2,2;\frac{sM}{2\mu _KT}),
\eeq
where $\Psi (a,b;z)$ is the confluent hypergeometric function [32].

The formula (2.4) should be considered as the analytic continuation of
(2.6). As a power series, (2.6) diverges for $|e^{\mu /T}|\geq 1,$
i.e., for $\mu \geq 0,$ while (2.4) is always valid.

In the limit $\frac{M}{\mu _KT}>>1$ (or $T<<\frac{M}{\mu _K}),$ Eq.
(2.6) reduces to [27]
\beq
n=\pm \frac{1}{2\pi ^3}T^4Li_4(e^{\pm \frac{\mu }{T}}),
\eeq
where $Li_n(z)\equiv \sum _{k=1}^\infty \frac{z^k}{k^n}$ is the
polylogarithm [33]. One can also obtain in this limit,
\bqry
p & = & \pm \frac{2}{\pi ^3}\frac{T_{\triangle V}}{M}T^6Li_6(\pm
e^{\frac{\mu }{T}}), \NL
\rho & = & \pm \frac{10}{\pi ^3}\frac{T_{\triangle V}}{M}T^6Li_6(\pm
e^{\frac{\mu }{T}})\;=\;5p, \NL
N_0 & = & \pm \frac{2}{\pi ^3}\frac{T_{\triangle V}}{M}T^5Li_5(\pm e^{
\frac{\mu }{T}}),
\eqry
where $p$ is the pressure and $\rho $ is the density of energy of the
$particle$ gas, and $N_0$ is the density of $particles$ per unit space
volume; $T_{\triangle V}$ is the average passage
interval in $\tau $ for the events which pass through a small (typical)
four-volume $\triangle V$ in the neighborhood of the $R^4$-point [25].

\section{Introduction of antiparticles}
In establishing the theory presented in this paper we assumed that the
total number of events and, therefore, particles, is a conserved
quantity, so it makes sense to talk of a box of $N$ particles. This can
no longer be true at high temperatures [34]; it is well known that at
temperatures $T\stackrel{>}{\sim }m$ quantum field theory requires the
inclusion of particle-antiparticle pair production, which becomes
important and therefore should be taken into account.
If $\bar{N}$ is the number of antiparticles, then $N$ and
$\bar{N}$ by themselves are not conserved but $N-\bar{N}$ is.
Therefore, the high-temperature limit of (2.4) is not relevant in
realistic physical systems.

The introduction of antiparticles into the theory in a systematic way was
made by Haber and Weldon [34] within the framework of the usual on-shell
theory. They considered an ideal Bose gas
with a conserved quantum number (referred to as ``charge'') $Q,$ which
corresponds to a quantum mechanical particle number operator commuting
with the Hamiltonian $\hat{H}.$ All thermodynamic quantities may be then
obtained from the grand partition function $Tr\;\{\exp [-(\hat{H}-\hat{Q}
)/k_BT]\}$ considered as a function of $T,\;V,$ and $\mu $ \cite{Hua}.
The formula for the conserved net charge
reads\footnote{One uses the standard recipe according to which all
additive thermodynamic quantities are reversed for antiparticles.} [34]
\beq
Q=\sum _{{\bf k}}\left[\frac{1}{e^{(E_k-\mu )/k_BT}-1}-
\frac{1}{e^{(E_k+\mu )/k_BT}-1}\right].
\eeq
In such a formulation a boson-antiboson system is described by only one
chemical potential $\mu ;$ the sign of $\mu $ indicates whether particles
outnumber antiparticles or vice versa. For the case of fermions one
has simply to change the sign in the denominators of Eq. (3.1).

The introduction of antiparticles into the theory we are discussing here,
as the events in the $CPT$-conjugate state, having nonnegative energy,
leads, by application of the arguments of Haber and Weldon, to
a change in sign of $\mu $ in the distribution function for
antiparticles. A full consideration of particle-antiparticle pair
production within the off-shell framework should include
$anti$-$events$ as well, i.e.,
events having the opposite sign of $\mu _K.$

The full theory of anti-events is constructed in \cite{prep}.
Upon $\tau $ reversal combined by charge conjugation, they are
included in statistical mechanics as the usual events (with positive
$M),$ having the opposite sign of $\mu _K$ in the distribution function
\cite{prep},[37]. The following relation which generalizes (2.4)
is found as the analogue of the formula (3.1):
\bqry
N & = & N_E+\bar{N}_{\bar{E}}-N_{\bar{E}}-\bar{N}_{E}
\NL
& = & N(\mu ,\mu _K)+N(-\mu ,-\mu _K)
-N(-\mu ,\mu _K)-N(\mu ,-\mu _K) \NL
& = & \sum _{k^\mu }[n_{k^\mu }(\mu ,\mu _K)+n_{k^\mu }(-\mu ,-\mu _K)
-n_{k^\mu }(-\mu ,\mu _K)-n_{k^\mu }(\mu ,-\mu _K)] \NL
& = & \sum _{k^\mu }\left[\frac{1}{e^{(E-\mu +\mu _K\frac{m^2}{2M})/k_BT}
\mp 1}+\frac{1}{e^{(E+\mu -\mu _K\frac{m^2}{2M})/k_BT}\mp 1}\right. \NL
 &   & \left. \;-\;\frac{1}{e^{(E+\mu +\mu _K\frac{m^2}{2M})/k_BT}\mp 1}
-\frac{1}{e^{(E-\mu -\mu _K\frac{m^2}{2M})/k_BT}\mp 1}\right]
\eqry
where $N$ is the conserved net event ``charge'', $N_E$ and
$N_{\bar{E}}$ are the numbers of events with $E\geq 0,\;E<0,$
respectively, and $\bar{N}_E,\;\bar{N}_{\bar{E}}$ are the corresponding
numbers of anti-events.

Replacing summation by integration, we now obtain a similar
formula for the event number densities:
\bqry
n & = & n(\mu ,\mu _K)+n(-\mu ,-\mu _K)-n(-\mu ,\mu _K)-
n(\mu ,-\mu _K) \NL
  & = & \frac{1}{4\pi ^3}\int _0^\infty m^3dm\sinh ^2\beta d\beta \left[
\frac{1}{e^{(m\cosh \beta -\mu +\mu _K\frac{m^2}{2M})/T}\mp 1}+
\frac{1}{e^{(m\cosh \beta +\mu -\mu _K\frac{m^2}{2M})/T}\mp 1}\right. \NL
&  & \;\;\;\;\;\;\;\;\;\;\;\;\;\;\;\;\;\;\;\;\;\;\;\;\;\;\;\;\;\;\;\;\;\;
\;\;\left. -\;\frac{1}{e^{(m\cosh \beta +\mu +\mu _K\frac{m^2}{2M})/T}\mp
1}-\frac{1}{e^{(m\cosh \beta -\mu -\mu _K\frac{m^2}{2M})/T}\mp 1}
\right]. \NL
 &  &
\eqry
The event--anti-event system possessing Bose-Einstein statistics was
treated in ref. [37]. The requirement that the four $n_{k^\mu }$'s in Eq.
(3.2) be positive definite leads to the bounds on the mass spectrum [37]
\beq
\frac{M}{|\mu _K|}\left(1-\sqrt{1-\frac{2|\mu \mu _K|}{M}}\right)\leq m
\leq \frac{M}{|\mu _K|}\left(1+\sqrt{1-\frac{2|\mu \mu _K|}{M}}\right),
\eeq
and to the relation
\beq
|\mu \mu _K|\leq \frac{M}{2}.
\eeq
The mass region (3.4) for small $\mu $ is well approximated by
\beq
|\mu |\leq m\leq \frac{2M}{|\mu _K|}.
\eeq
Therefore, for $|\mu |\simeq 0$ and $|\mu _K|\rightarrow 0$ we have
almost the whole range of $m,$ $(0,\infty ).$ Since there is no
requirement on the $n_{k^\mu }$'s in the fermionic case, we take for
this case $0\leq m<\infty .$ Thus, both cases can be treated
simultaneously, with the whole range of $m.$ In this way we obtain, upon
expansion of the denominators in Eq. (3.3) into power series and
integration on $\beta ,$
\beq
n=-\frac{T}{\pi ^3}\int _0^\infty dm\sum _{s=1}^\infty \frac{(\pm 1)^{s+
1}}{s}m^2\sinh \frac{s\mu }{T}\sinh \frac{s\mu _Km^2}{2MT}K_1\left(\frac{
sm}{T}\right),
\eeq
from which we identify the mass distribution $f(m),$ normalized
as\footnote{We remark that only $|n|$ has physical meaning as a
conserved net event charge.}
\beq
\Big|\int _0^\infty dmf(m)\Big|=|n|.
\eeq
In (3.7), $K_1$ is a Bessel function of the third kind (imaginary
argument).

Integration in (3.7) gives the following expression for $n:$
\beq
n=-\frac{1}{4\pi ^3}\frac{M^2}{\mu _K^2}T^2\sum _{s=1}^\infty \frac{(\pm
1)^{s+1}}{s^2}\sinh \frac{s\mu }{T}\left[\Psi (2,2;\frac{sM}{2\mu _KT})-
\Psi (2,2;-\frac{sM}{2\mu _KT})\right].
\eeq
The formulas for $p,\rho $ and $N_0$ become \cite{prep}
\bqry
p & = & p(\mu ,\mu _K)+p(-\mu ,-\mu _K)+p(-\mu ,\mu _K)+p(\mu
,-\mu _K), \\
\rho  & = & \rho (\mu ,\mu _K)+\rho (-\mu ,-\mu _K)+
\rho (-\mu ,\mu _K)+\rho (\mu ,-\mu _K), \\
N_0 & = & N_0(\mu ,\mu _K)+N_0(\mu ,-\mu _K)-
N_0(-\mu ,\mu _K)-N_0(-\mu ,-\mu _K),
\eqry
where $p(\mu ,\mu _K),\;\rho (\mu ,\mu _K)$ and $N_0(\mu ,\mu _K)$ are
the subdistributions defined by the corresponding formulas of the event
statistical mechanics [27].

Note that the expressions (3.12) for $N_0$ (actually corresponding to the
$Q$ in Eq. (3.1)), which refers to
particles and antiparticles, and (3.3) for $n,$ which refers to
{\it events,} differ in sign of the corresponding terms
containing $-\mu _K.$ While $N_0$ commutes with the effective
Hamiltonian of a relativistic many body system in the usual framework,
it is (3.2) which commutes with the evolution operator of the covariant
theory \cite{prep}.

We also remark that, as in the usual theory, the system we are studying
is described by only one chemical potential $\mu $ and only one mass
potential $\mu _K;$ the sign of $\mu $ indicates whether particles
outnumber antiparticles or vice versa, while the sign of $\mu \mu _K$
has a similar relation to the relative number of
events and anti-events, as seen in Eqs. (4.8).

\section{Realistic equation of state for event--anti-event system}
In the limit $\frac{M}{|\mu _K|}>>T,$ we use the asymptotic formula for
$z\rightarrow \infty $ (ref. [32], p.278, subsection 6.13.1)
\beq
\Psi (a,a;z)\sim z^{-a}\left(1-\frac{a}{z}+\frac{a(a+1)}{z^2}-
\frac{a(a+1)(a+2)}{z^3}+\ldots \right),
\eeq
and obtain from (3.9), retaining the terms up to the order of $(\frac{
\mu _KT}{M})^2,$
\bqry
n & = & \mp \frac{4}{\pi ^3}\frac{\mu _K}{M}T^5\left.\Big\{\left[Li_5(\pm
e^{\frac{\mu }{T}})-Li_5(\pm e^{-\frac{\mu }{T}})\right]\right. \NL
 &  & \left. +48\frac{\mu _K^2T^2}{M^2}\left[Li_7(\pm e^{\frac{\mu }
{T}})-Li_7(\pm e^{-\frac{\mu }{T}})\right]\right\}. \\
\eqry
The corresponding formulas for $p,$ $\rho $ and $N_0$ read [27]
\bqry
p & = & \pm \frac{4}{\pi ^3}\frac{T_{\triangle V}}{M}T^6\left. \Big\{
\left[Li_6(\pm e^{\frac{\mu }{T}})+Li_6(\pm e^{-\frac{\mu }{T}})\right]
\right. \NL
 &  & \left. +48\frac{\mu _K^2T^2}{M^2}\left[Li_8(\pm e^{\frac{\mu
}{T}})+Li_8(\pm e^{-\frac{\mu }{T}})\right]\right\}, \\
\rho & = & \pm \frac{4}{\pi ^3}\frac{T_{\triangle V}}{M}T^2\left.\Big\{ 5
\left[Li_6(\pm e^{\frac{\mu }{T}})+Li_6(\pm e^{-\frac{\mu
}{T}})\right]\right. \NL
 &  & \left. +432\frac{\mu _K^2T^2}{M^2}\left[Li_8(\pm e^{\frac{\mu
}{T}})+Li_8(\pm e^{-\frac{\mu }{T}})\right]\right\}, \\
N_0 & = & \pm \frac{4}{\pi ^3}\frac{T_{\triangle V}}{M}T^5\left. \Big\{
\left[Li_5(\pm e^{\frac{\mu }{T}})-Li_5(\pm e^{-\frac{\mu
}{T}})\right]\right. \NL
 &  & \left. +48\frac{\mu _K^2T^2}{M^2}\left[Li_7(\pm e^{\frac{\mu
}{T}})-Li_7(\pm e^{-\frac{\mu }{T}})\right]\right\}.
\eqry
Note the remarkable relation
\beq
\Big|N_0\Big|=\frac{T_{\triangle V}}{|\mu _K|}|n|.
\eeq
First consider the bosonic case. Using the corresponding relations for
the polylogarithms (see Appendix A), and retaining the terms up to the
order of $(\frac{\mu _KT}{M})^2,$ $(\frac{\mu }{T})^2,$ we obtain
\bqry
n & = & -\frac{4\pi }{45}\frac{\mu \mu _K}{M}
T^4\left(1+\frac{5}{2\pi ^2}\frac{\mu ^2}{T^2}+
\frac{96\pi ^2}{21}\frac{\mu _K^2T^2}{M^2}\right), \NL
p & = & \;\frac{8\pi ^3}{945}\frac{T_{\triangle V}}{M}T^6\left(1+
\frac{21}{4\pi ^2}\frac{\mu ^2}{T^2}+\frac{24\pi ^2}{5}
\frac{\mu _K^2T^2}{M^2}\right), \NL
\rho  & = & \frac{40\pi ^3}{945}\frac{T_{\triangle V}}{M}T^6\left(1
+\frac{21}{4\pi ^2}\frac{\mu ^2}{T^2}+\frac{216\pi ^2}{25}
\frac{\mu _K^2T^2}{M^2}\right), \NL
N_0 & = & \;\frac{4\pi }{45}\frac{T_{\triangle V}}{M}\mu T^4\left(1+
\frac{5}{2\pi ^2}\frac{\mu ^2}{T^2}+\frac{96\pi ^2}{21}
\frac{\mu _K^2T^2}{M^2}\right).
\eqry
Therefore,
\beq
c^2=\frac{dp}{d\rho }\simeq \frac{p}{\rho }\cong \frac{1}{5}\left(1-
\frac{96}{25}\frac{\mu _K^2T^2}{M^2}\right),
\eeq
i.e., one obtains the value of the velocity of sound $c^2=0.20,$ as well
as the correction to this value, of the order of
$(\frac{\mu _KT}{M})^2.$

In the fermionic case, one uses the corresponding relations for the
polylogarithms with the argument $z<0$ (Appendix A), and obtains similar
formulas for $n,$ $p,$ $\rho $ and $N_0,$ which give
$$c^2\cong \frac{1}{5}\left(1-\frac{127}{124}
\;\frac{96}{25}\;\frac{\mu _K^2(k_BT)^2}{M^2}\right).$$
We see that in the case $\mu \simeq 0,$ $\mu _K\rightarrow 0$ the
equation of state for event--anti-event system reads
$p,\rho \propto T^6,$ implying the velocity of sound $c^2=0.20,$ in
agreement with the Shuryak's ``realistic'' equation of state. We now turn
to the corresponding distribution of mass.

\section{Relativistic mass distribution in event--anti-event system}
The distribution of mass, defined by Eq. (3.7), can be simplified in the
case $\mu \simeq 0,$ $\frac{M}{|\mu _K|T}>>1.$ Since for the whole range
of $m$ and small $\mu $ the average $\langle m\rangle $ depends only on
temperature [26],[27], $\frac{|\mu _K|m^2}{2MT}$ is of the order of
$\frac{|\mu _K|T}{M}<<1.$ Therefore, one can replace the hyperbolic sine
functions by their arguments and obtain in this way
\beq
f(m)\simeq \frac{1}{2\pi ^3}\frac{|\mu \mu _K|}{MT}\sum _{s=1}^\infty
(\pm 1)^{s+1}sm^4K_1\left(\frac{sm}{T}\right).
\eeq
For a given $T,$ in the region $m<<T$ we use the asymptotic formula \\
(ref. \cite{AS}, p.375) $$K_\nu (z)\sim \frac{1}{2}\Gamma (\nu
)\left(\frac{z}{2}\right)^{-\nu },\;\;\;z\rightarrow 0,$$ and obtain
\beq
f(m)\sim m^3,
\eeq
for both bosonic and fermionic cases. If we assume that the distribution
(5.2) describes the genuine spectrum of compound hadron-resonance
systems, and restrict ourselves only to the region $m<<T,$ we shall
obtain, with Shuryak's Eqs. (1.3),(1.8) for $k=3,$ $$c^2=0.14.$$ In the
region $m>>T,$ we use another asymptotic formula (ref. \cite{AS}, p. 378)
$$K_\nu (z)\sim \sqrt{\frac{\pi }{2z}}e^{-z},\;\;\;z\rightarrow
\infty .$$ Since in this region particles become distinguishable, we have
\beq
f(m)\sim m^{7/2}e^{-\frac{m}{T}}.
\eeq
In this case the equation of state tends asymptotically to the equation
of state of an ideal Stefan-Boltzmann gas, $p=\rho /3,$ providing for the
velocity of sound the value $$c^2=0.33.$$ For the transitional region
$m\stackrel{<}{\sim }T,$ the formula for $K_\nu (z)$ can be found by the
method of steepest descents (ref. [32], Vol.2, p.28). The corresponding
expression for the mass distribution can be also obtained directly from
Eq. (5.3), since in this case the power function dominates over the
exponent. In either case, one has
\beq
f(m)\sim m^{7/2},
\eeq
so that, if we restrict ourselves only to this region, we shall obtain,
through Eqs. (1.3),(1.8) with $k=7/2,$ $$c^2=\frac{2}{15}\approx 0.13.$$
If the three cases are treated on an equal footing, one has
\beq
c^2=\frac{1}{3}(0.14+0.13+0.33)=0.20,
\eeq
in agreement with the equation of state (4.9) for the whole mass region.
We see that the mass distribution (5.1) is a good candidate for
fitting the hadronic resonances, in good agreement with the
phenomenological models of Hagedorn, Shuryak, {\it et al.}

\subsection{Specific heat of event--anti-event gas}
In conclusion we wish to calculate the specific heat of the
event--anti-event gas we are studying. The distribution (5.1) gives
the following value of the average $m:$
\beq
\langle m\rangle =\frac{\int dm\;mf(m)}{\int dmf(m)}=\frac{\int _0^\infty
dm\;m^5K_1(\frac{m}{T})}{\int _0^\infty dm\;m^4K_1(\frac{m}{T})}=\frac{45
\pi }{32}T,
\eeq
where we used the formula (ref. [39], p.684, formula 16)
\beq
\int _0^\infty dx\;x^\mu K_\nu (ax)=2^{\mu -1}a^{-\mu -1}\Gamma \left(
\frac{1+\mu +\nu }{2}\right)\Gamma \left(\frac{1+\mu -\nu }{2}\right).
\eeq
The rest frame average of the energy $\langle m\cosh \beta \rangle $
can be obtained from (3.3),(5.1), with the help of the relation $$\sinh
^2\beta \cosh \beta =\frac{1}{4}(\cosh 3\beta -\cosh \beta )$$ and the
formula (ref. [39], p.358, formula 4)
\beq
\int _0^\infty dx\cosh \nu x\;e^{-a\cosh x}=K_\nu (a),
\eeq
as follows,
\bqry
\langle E\rangle  & = & \frac{1}{n}\int m^3dm\sinh ^2\beta d\beta \left[
\frac{m\cosh \beta }{e^{(m\cosh \beta -\mu +\mu _K\frac{m^2}{2M})/T}\mp 1
}\;+\;{\rm three\;\;terms}\;\;\right] \NL
& = & \frac{1}{4T}\frac{\int _0^\infty dm\;m^6\left[K_3(\frac{m}{T})-K_1(
\frac{m}{T})\right]}{\int _0^\infty dm\;m^4K_1(\frac{m}{T})}=6T.
\eqry
This value is twice as much as Pauli's classical value $3T$ obtained
in the ultrarelativistic limit within the usual on-shell theory \cite{P},
which is quite natural, since we take account of particle-antiparticle
pairs. The value of the specific heat is now obtained from (5.6),(5.9):
\beq
\langle E-m\rangle =\left(6-\frac{45\pi }{32}\right)T=\gamma ^{'}\frac{3
}{2}T,
\eeq
\beq
\gamma ^{'}=4-\frac{15\pi }{16}\approx 1.055.
\eeq
We see that at low temperatures, $\gamma ^{'}$ provides a correction of
the order of 5.5\% to the classical value $3/2.$

\section{Concluding remarks}
We have discussed possible consequences of a manifestly covariant
relativistic statistical mechanics for hadronic physics. We have
considered the relativistic event--anti-event system, for which we have
obtained the relativistic mass distribution and calculated all
characteristic thermodynamic variables. We have shown that in the case of
small $\mu _K$ and zero net particle charge, the equation of state for
such a system corresponds to the ``realistic'' equation of
state proposed by Shuryak [4] to describe the behavior of hot hadronic
matter. We have seen that the mass distribution obtained in this paper
can serve as a framework for fitting the hadronic resonances, in good
agreement with the phenomenological models of Hagedorn {\it et al}
[8]-[10].

In ref. [41] we have argued that the mass potential $\mu _K$ is
related to the degeneracy parameter $\delta :$ $\mu _K\sim \delta ^{-a},
\;a>0.$ Therefore, small $\mu _K$ indicates large $\delta ,$ i.e., high
degeneracy. In this case the system should exhibit the collective
behavior determined by a relatively strong interaction, mainly of
attractive character. This attraction considerably reduces the pressure,
resulting in a value for the sound velocity $c^2\approx 0.20.$

Further physical consequences of the theory of the off-shell events are
discussed in refs. \cite{cond},[41].
In \cite{cond} the statistical mechanics of the bosonic
event--anti-event system is considered. For such a system, at some
critical temperature a special type of Bose-Einstein condensation sets
in, which provides the events making up the ensemble a definite mass and
represents, in this way, a phase transition to an equilibrium on-shell
sector. In [12] an adiabatic equation of state, $p\propto N_0^{6/5},$ is
obtained for the system of degenerate off-shell fermions and possible
implications in astrophysics are discussed.

The various aspects of the theory of the off-shell events discussed in
this paper, as well as related questions, are now under further
investigation.

\newpage
\section*{Appendix A}
\setcounter{equation}{0}
\renewcommand{\theequation}{A.\arabic{equation}}
The polylogarithm is defined by the relation
\beq
Li_\nu (z)=\sum _{k=1}^\infty \frac{z^k}{k^\nu },\;\;\;|z|<1\;\;{\rm or}
\;\;|z|=1,\;\;Re\;\nu >1.
\eeq
For $|z|\geq 1,$ the function $Li_\nu (z)$ is defined as the analytic
continuation of this series.

For integer $n$ and positive $z,$
the following relation holds (ref. [33], p.763):
\beq
Li_n(z)+(-1)^nLi_n\left(\frac{1}{z}\right)=-\frac{(2\pi i)^n}{n!}B_n
\left(\frac{\ln z}{2\pi i}\right),
\eeq
where $B_n$ are the Bernoulli polynomials (ref. [33], p.765),
$$B_0(z)=1,\;\;\;B_1(z)=z-\frac{1}{2},\;\;\;B_2(z)=z^2-z+\frac{1}{6},$$
$$B_3(z)=z^3-\frac{3}{2}z^2+\frac{1}{2}z,\;\;\;B_4(z)=z^4-2z^3+z^2-\frac{
1}{30},\;\;\;{\rm etc.}$$
possessing the property $$B_n^{'}(z)=nB_{n-1}(z).$$
In view of (A.2) we have
\bqry
Li_1(z)-Li_1\left(\frac{1}{z}\right) & = & -\ln z+\pi i, \\
Li_2(z)+Li_2\left(\frac{1}{z}\right) & = &
-\frac{\ln ^2z}{2}+\pi i\ln z+\frac{\pi ^2}{3}, \\
Li_3(z)-Li_3\left(\frac{1}{z}\right) & = &
-\frac{\ln ^3z}{6}+\frac{\pi i}{2}\ln ^2z+\frac{\pi ^2}{3}\ln z, \\
Li_4(z)+Li_4\left(\frac{1}{z}\right) & = & -\frac{\ln ^4z}{24}+\frac{
\pi i}{6}\ln ^3z+\frac{\pi ^2}{6}\ln ^2z+\frac{\pi ^4}{45}, \\
Li_5(z)-Li_5\left(\frac{1}{z}\right) & = & -\frac{\ln ^5z}{120}+\frac{
\pi i}{24}\ln ^4z+\frac{\pi ^2}{18}\ln ^3z+\frac{\pi ^4}{45}\ln z, \\
Li_6(z)+Li_6\left(\frac{1}{z}\right) & = &
-\frac{\ln ^6z}{720}+\frac{\pi i}{120}\ln ^5z\!+\!\frac{\pi ^2}{72}
\ln ^4z\!+\!\frac{\pi ^4}{90}\ln ^2z\!+\!\frac{2\pi ^6}{945},\;{\rm etc.}
\eqry
\newpage
For integer $n$ and negative $z,$  we use another relation
(ref. [33], p.763):
\beq
Li_n(z)+(-1)^nLi_n\left(\frac{1}{z}\right)=-\sum _{k=1}^{[n/2]}c_{k,n}
\ln ^{n-2k}(-z);
\eeq
$$c_{0,n}=\frac{1}{n!},\;\;\;c_{1,n}=\frac{\pi ^2}{6(n-2)!},$$
$$c_{k,n}=(-1)^{k+1}\frac{2(2^{2k-1}-1)\pi ^{2k}B_{2k}}{(2k)!(n-2k)!},\;
\;\;k=0,1,\ldots ,[n/2],$$ where $B_{2k}$ are the Bernoulli numbers,
$$B_0=1,\;\;B_2=\frac{1}{6},\;\;B_4=-\frac{1}{30},\;\;B_6=\frac{1}{42},\;
\;{\rm etc.}$$ In this way we obtain
\bqry
Li_1(z)-Li_1\left(\frac{1}{z}\right) & = & -\ln (-z), \\
Li_2(z)+Li_2\left(\frac{1}{z}\right) & = &
-\left[\frac{\ln ^2(-z)}{2}+\frac{\pi ^2}{6}\right], \\
Li_3(z)-Li_3\left(\frac{1}{z}\right) & = &
-\left[\frac{\ln ^3(-z)}{6}+\frac{\pi ^2}{6}\ln (-z)\right], \\
Li_4(z)+Li_4\left(\frac{1}{z}\right) & = & -\left[\frac{\ln ^4(-z)}{24}
+\frac{\pi ^2}{12}\ln ^2(-z)+\frac{7\pi ^4}{360}\right], \\
Li_5(z)-Li_5\left(\frac{1}{z}\right) & = & -\left[\frac{\ln ^5(-z)}{120}
+\frac{\pi ^2}{36}\ln ^3(-z)+\frac{7\pi ^4}{360}\ln (-z)\right], \\
Li_6(z)+Li_6\left(\frac{1}{z}\right) & = &
-\left[\frac{\ln ^6(-z)}{720}\!+\!\frac{\pi ^2}{144}\ln ^4(-z)\!+\!
\frac{7\pi ^4}{720}\ln ^2(-z)\!+\!\frac{31\pi ^6}{15120}\!\right]
\eqry
$\;\;\;\;\;\;\;\;\;\;\;\;\;\;\;\;\;\;\;\;\;\;\;\;
\;\;\;\;\;\;\;\;\;\;\;\;\;\;\;\;\;\;\;\;$ etc. \\
The formulas in the main text are obtained from (A.3)-(A.8) for $z=e^{
\frac{\mu }{T}},$ and from (A.10)-(A.15) for $z=-e^{\frac{\mu }{T}},$ for
the bosonic and the fermionic cases, respectively.

\end{document}